\documentclass{article}

\usepackage[english]{babel}

\usepackage[a4paper,top=2cm,bottom=2cm,left=3cm,right=3cm,marginparwidth=1.75cm]{geometry}

\usepackage{amsmath}
\usepackage{graphicx}
\usepackage{authblk}                      

\usepackage[colorlinks=true, allcolors=blue]{hyperref}

\title{Results from ON-OFF analysis of the Neutrinos-Angra detector}

\author[1,*]{E. Kemp}
\author[1]{W. V. Santos }
\author[2]{J. C. Anjos}
\author[3]{P. Chimenti}
\author[4]{L. F. G. Gonzalez}
\author[5]{G. P. Guedes}
\author[2,6]{H. P. Lima Jr.}
\author[7]{R. A. Nóbrega}
\author[8]{I. M. Pepe}
\author[8]{D. B. S. Ribeiro}

\affil[1]{\footnotesize Instituto de Física Gleb Wataghin, Universidade Estadual de Campinas, Campinas, 13083-859, SP, Brazil}
\affil[2]{\footnotesize Centro Brasileiro de Pesquisas Físicas, Rio de Janeiro, 22290-180, RJ, Brazil}
\affil[3]{\footnotesize Departamento de Física, Universidade Estadual de Londrina, Londrina, 86057-970, PR, Brazil}
\affil[4]{\footnotesize Instituto de Computação, Universidade Estadual de Campinas, Campinas, 13083-852, SP, Brazil}
\affil[5]{\footnotesize Universidade Estadual de Feira de Santana, Feira de Santana, 44036-900, BA, Brazil}
\affil[7]{\footnotesize Astroparticle Physics Division, Gran Sasso Science Institute, L’Aquila, 67100, Italy}
\affil[7]{\footnotesize Departamento de Circuitos Elétricos, Universidade Federal de Juiz de Fora, Juiz de Fora, 36036-900, MG, Brazil}
\affil[8]{\footnotesize Instituto de Física, Universidade Federal da Bahia, Salvador, 40170-115, BA, Brazil}

\affil[*]{\footnotesize\textbf{Corresponding author: 
 } \href{mailto:kemp@unicamp.br}{kemp@unicamp.br}}

\date{June 26, 2024}

\begin{document}
\maketitle

\begin{abstract}
The Neutrinos Angra Experiment, a water-based Cherenkov detector, is located at the Angra dos Reis nuclear power plant in Brazil.  Designed to detect electron antineutrinos produced in the nuclear reactor, the primary objective of the experiment is to demonstrate the feasibility of monitoring reactor activity using an antineutrino detector. This effort aligns with the International Atomic Energy Agency (IAEA) program to identify potential and novel technologies applicable to nonproliferation safeguards.

Operating on the surface presents challenges such as high noise rates, necessitating the development of very sensitive, yet small-scale detectors. These conditions make the Angra experiment an excellent platform for both developing the application and gaining expertise in new technologies and analysis methods. The detector employs a water-based target doped with gadolinium to enhance its sensitivity to antineutrinos.

In this work, we describe the main features of the detector and the electronics chain, including front-end and data acquisition components. We detail the data acquisition strategies and the methodologies applied for signal processing and event selection. Preliminary physics results suggest that the detector can reliably monitor reactor operations by detecting the inverse beta decay induced by electron antineutrinos from the reactor.
\end{abstract}


\section{Introduction}\label{intro}

The $\nu$-Angra experiment aims to develop an antineutrino detector designed to improve nuclear safety at the Angra dos Reis power plant in Brazil. Nuclear reactors have been crucial to experimental neutrino physics, as they are numerous man-made sources of neutrinos. The detection of antineutrinos produced by nuclear power plants was first achieved in 1956 by pioneers C.L. Cowan and F. Reines, who detected the Inverse Beta Decay reaction (IBD): $\Bar{\nu_e} + p \rightarrow n + e^{+}$ \cite{cowan}. Since then, other experiments have been built inside such facilities, with reactor neutrino experiments significantly contributing to the precise knowledge of the neutrino oscillation parameters \cite{kamland, dchooz, dbay, reno}.

The proposal of using neutrinos for remote monitoring of nuclear reactor thermal power was first considered in the mid-1970s \cite{Mik, BoroMik}. One of the first demonstration experiments was performed in a neutrino laboratory located in the Rovno nuclear power plant, Ukraine, where the relationship between neutrino count rate and reactor activity was clearly shown \cite{Rovno1, Rovno2}. This scenario opens up solid perspectives for using neutrinos as reliable probes of the physical processes in which they participate, making a neutrino detector capable of monitoring parameters related to the activity of nuclear reactors crucial for non-proliferation safeguards dictated by the International Atomic Energy Agency (IAEA) \cite{Bernstein:2019hix, korea}.

The $\nu$-Angra experiment employs a water Cherenkov detector that utilizes a non-flammable target positioned at the surface level. This experiment, initially considered for studies of neutrino oscillations, evolved to focus on the verification of nonproliferation safeguards, given the opportunity to conduct an experiment at the Angra dos Reis nuclear complex. The Angra-II reactor, in steady-state operation, provides a neutrino flux estimated as $1.21\times10^{20}~s^{-1}$ \cite{connie}. The electron antineutrino flux can be used to perform noninvasive monitoring of reactor activity and estimate the thermal power produced at the reactor core.

The long-term goal of the $\nu$-Angra experiment is to develop a reliable and cost-effective technology to routinely monitor nuclear reactor power and possibly neutrino spectral evolution, which can reveal the composition of burned fuel, with special interest in the fraction of plutonium \cite{hubber, safeguards}. The stability of data acquisition is a fundamental step in this direction, and in the commissioning phase, we have fully validated the electronics that operate within the desired stability for years of data collection \cite{Lima:2018spe}.

A common challenge for all neutrino experiments is the background radiation, mainly neutrons and cosmic muons, that can mimic signals expected from neutrino detection. To suppress background levels, detectors are usually installed in large underground caverns using rock and soil overburdens as natural shields. However, the $\nu$-Angra detector is assembled on the surface, adhering to the agreements with the plant operator, which presents a significant challenge in handling an extremely low signal-to-noise ratio with enough ability to find genuine IBD events to monitor the Angra II nuclear reactor.

This article is structured as follows. Initially, we describe the water Cherenkov detector located near the Angra-II power plant. Next, we present the data acquisition system. Subsequently, we explore the data analysis phase, which encompasses pre-processing, data selection, and the detection of neutrino interactions. Finally, we conclude with a discussion of our findings and the implications of our research.


\section{The $\nu$-Angra detector}\label{detector}

Constructed in the 1980s near Angra dos Reis (RJ), the Angra-II nuclear power plant operates with a nominal thermal power of 3.8 GW. The Neutrino Physics Laboratory, housed within a standard high-cube container, is strategically placed about 25 meters from the reactor core, just outside and close to the containment dome of Angra-II. In contrast, Angra-I, another reactor core at the plant, is located approximately 200 meters away, whose contribution to the neutrino flux can be disregarded in the current context. Currently, this site hosts two significant experiments: $\nu$-Angra and CONNIE (Coherent Neutrino-Nucleus Interaction Experiment) \cite{CONNIE:2019swq}.

\begin{figure}[!ht]
\centering
\includegraphics[scale=0.4]{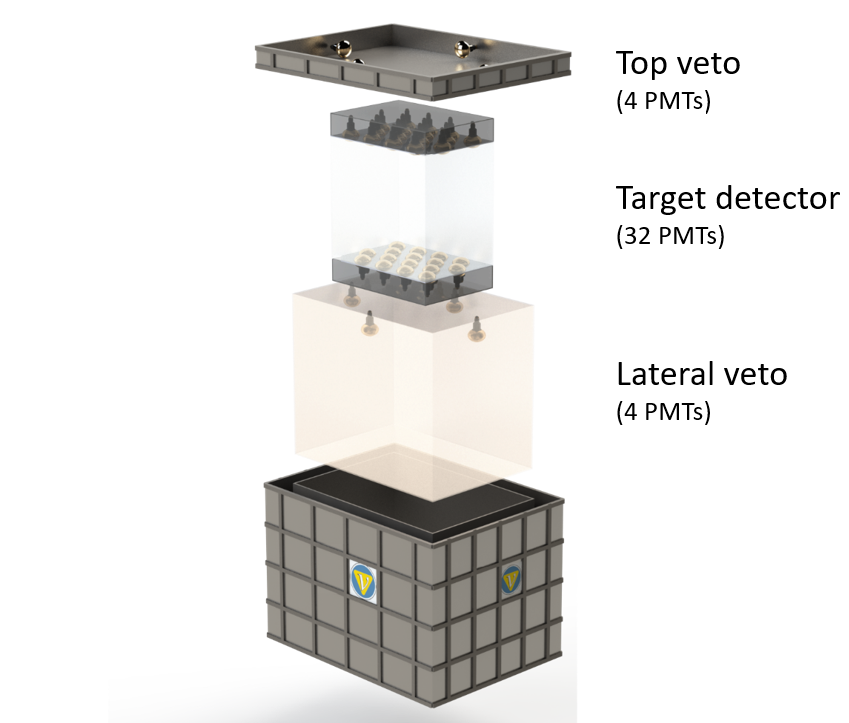}
\caption{Exploded view of the $\nu$-Angra detector.}
\label{fig:detector}
\end{figure}

The $\nu$-Angra experiment aims to develop an antineutrino detector that works as a tool for nuclear safeguards, using a water Cherenkov detector to measure the flux of antineutrinos emitted by the reactor. The $\nu$-Angra detector comprises three nested volumes, each playing a different role in the detection and mitigation of background signals.  The system features a central parallelepipedal tank, known as the target, which holds $\sim 1.3$ kL of water doped with 0.2\% gadolinium chloride (GdCl3) by mass. The target is monitored by 32 8-inch Hamamatsu R5912 photomultipliers (PMT), with 16 installed at the bottom and 16 at the top of the volume, as shown in Figure \ref{fig:detector}. Furthermore, the internal faces of the target are lined with GORE\textregistered, a light-diffusing membrane that ensures a reflectance of more than 99.0\% of photons $\sim$ 400 nm wavelength.

The electron antineutrino flux generated by the reactor can be used to perform non-invasive monitoring of the reactor activity and estimate the thermal power produced at the reactor core. The expected rate for the 1-ton target detector is around 5000 events per day considering the 25 m distance from the reactor core. This enables the experiment to investigate the potential of antineutrino detection for safeguards applications.

In the IBD interaction, the electron antineutrino ($\Bar{\nu_e}$) interacts with a hydrogen nucleus (proton, $p$), producing a positron ($e^+$) and a neutron ($n$): $ \Bar{{\nu_e}} + p \rightarrow e^{+} + n$. The positron promptly produces a Cherenkov flash\footnote{In water electrons or positrons with kinetic energies exceeding 264 keV emit Cherenkov radiation, detectable by the PMTs in the target.} and also annihilates with an electron from the detector material, creating two ($\gamma$) photons releasing additional energy for the reaction of approximately 511 keV each: $e^{+} + e^{-} \rightarrow \gamma + \gamma$.  This initial burst of light from the positron is termed the 'prompt signal'. Subsequently, the neutron, after thermalization, is captured by a gadolinium nucleus \cite{Hagiwara:2018kmr} within about $12.3\;\mu\text{s}$ \cite{Gonzalez_D}, leading to the emission of photons from deexcitation of the Gd atom. Typically, the reaction n(Gd, Gd$^*$)$\gamma$ releases a total gamma energy of $\sim$ 8 MeV. This subsequent photon emission is known as the 'delay signal'. Together, these IBD interactions produce distinct signals separated by the 'prompt' to 'delay' specific times and energy signatures.

To mitigate background noise from cosmic and environmental radiation, the detector's target is encased within two veto subsystems: the top veto and the lateral veto. The top veto is positioned directly above the target, while the lateral veto encircles it. Both subsystems are filled with ultra-pure water and each is equipped with four PMTs. The inner surfaces of these subsystems are lined with Tyvek\textregistered, another reflective liner with an efficiency greater than 97.0\% of light at pertinent wavelengths. Activation of two PMTs simultaneously in either subsystem is interpreted as background particles interference, prompting the data acquisition system to stop recording for that event.

Additionally, the lateral veto subsystem is encased in a nonactive volume known as the 'shield,' which features two sides measuring 14.5 cm in thickness and two others 22.5 cm thick, all filled with water. This configuration significantly enhances the subsystem's ability to block neutrons from cosmic radiation or the environment. The principal components of the $\nu$-Angra detector are illustrated in Figure \ref{fig:detector}.

\section{Readout Electronics and Data Acquisition}

\begin{figure}
	\centering
 \includegraphics[scale=.8]{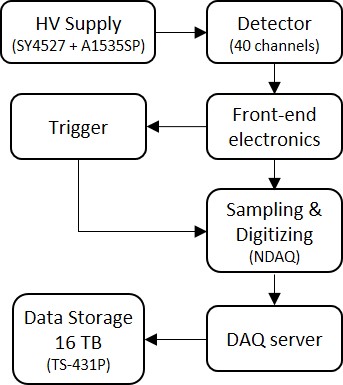}
	\caption{Overview of the complete electronics chain and readout scheme installed on the nuclear power plant laboratory.}
	\label{fig:daq_overview}
\end{figure}

The readout and data acquisition system comprises several key components: Front-End Electronics (FEE), Digitizers, and the Trigger. 
A diagram of the electronics and data acquisition chain is shown in Figure \ref{fig:daq_overview}.

\begin{figure}[!ht]

 \centering \includegraphics[width=.8\columnwidth]{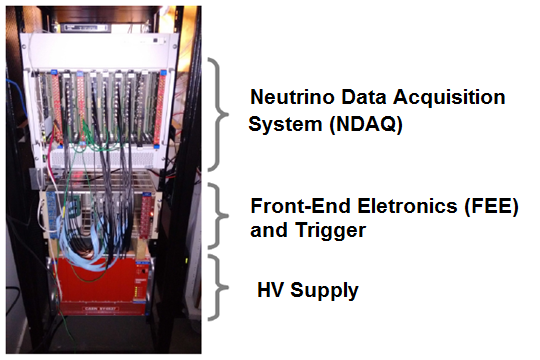}
 \caption{
   \label{fig:ElectronicRack} 
   Rack with $\nu$-Angra Detector Electronics \cite{Gonzalez_D}.
 }

\end{figure}

A comprehensive data acquisition system has been developed and implemented for the $\nu$-Angra experiment to fulfill several critical functions: biasing the PMTs with a high voltage power supply, amplifying the output signals from the PMTs via Front-End Electronics (FEE), sampling and digitizing the signals through the Neutrino Data Acquisition System (NDAQ), selecting events using the trigger system, and managing local data storage, which is mirrored at CBPF and Unicamp. Figure \ref{fig:ElectronicRack} shows a picture of the rack with the $\nu$-Angra electronics installed in our laboratory at the reactor site.

The HV power supply system, along with the local computing infrastructure (DAQ server) and data storage components, are commercial standard products.  The HV supply, which biases the 40 PMTs is a mainframe-based system manufactured by CAEN, and is remotely controlled via an Ethernet connection.  Local data storage comprises a network-connected storage unit that continuously records data from the detector. 
This data is subsequently transferred to two larger, permanent data servers: the primary server located at CBPF in Rio de Janeiro, and a secondary mirror server at Unicamp in Campinas.

\subsection{Front-End Electronics - FEE}

Custom front-end circuitry has been engineered for the $\nu$-Angra detector to process PMT output signals for digitization by NDAQ modules and to feed the trigger system with information from activated channels ~\cite{Dornelas:2016fee}.

\begin{figure}[ht!]
\centering
\includegraphics[scale=0.28]{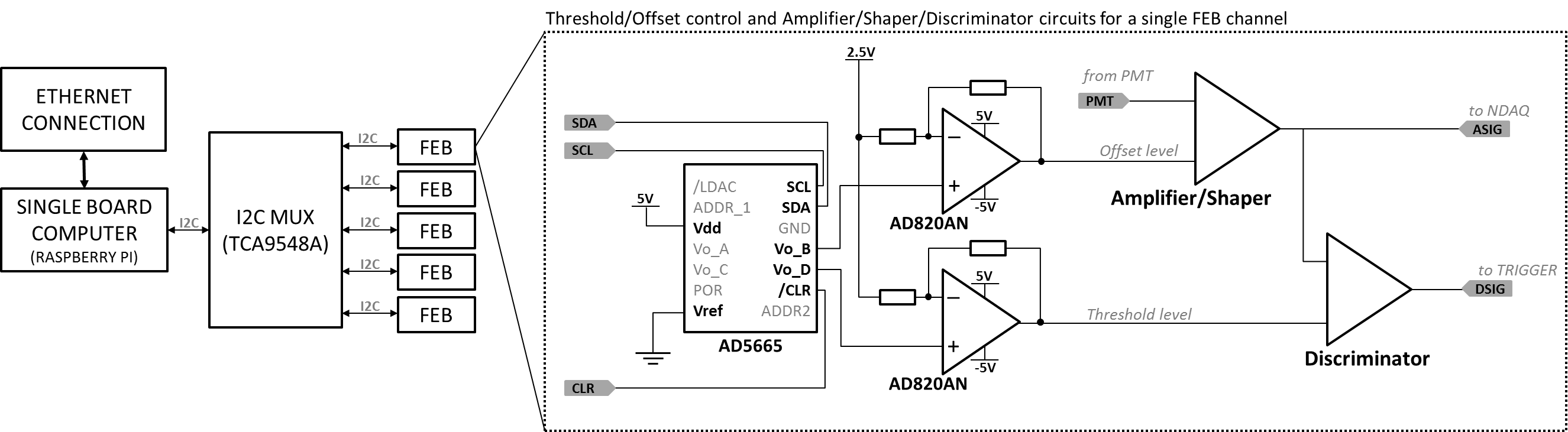}
\caption{Block diagram of 
 single FFE channel.}
\label{fig:fee_diag}
\end{figure}

Five front-end boards (FEBs), each containing eight independent channels, are integrated to the experiment. 
Each channel features a four-stage amplification/shaper circuit that conditions the signal for digitization, a discriminator circuit that relays outputs to the trigger system, and a control system. 
The control system employs a programmable Digital-to-Analog Converter (DAC) integrated circuit, enabling remote adjustments of the analog signal's offset and the discrimination thresholds according to experimental needs. The FEE functionality of each channel is depicted in a block diagram shown in Figure \ref{fig:fee_diag}.

\subsection{Digitizer}

The detector signals, amplified and shaped by the Front-End Electronics (FEE), are transmitted to the VME-based system that samples and digitizes them upon receiving a trigger pulse. The NDAQ board was specifically designed for the experiment~\cite{ndaq}. A VME single-board computer serves as the readout processor (ROP) for five NDAQ boards housed in a VME crate. The ROP manages and reads the NDAQ cards during the data acquisition phase. Additionally, a commercial fan-in fan-out module distributes the digital trigger pulse from the trigger system to the NDAQ modules, ensuring synchronized data handling.

\begin{figure}[ht!]
\centering
\includegraphics[scale=0.28]{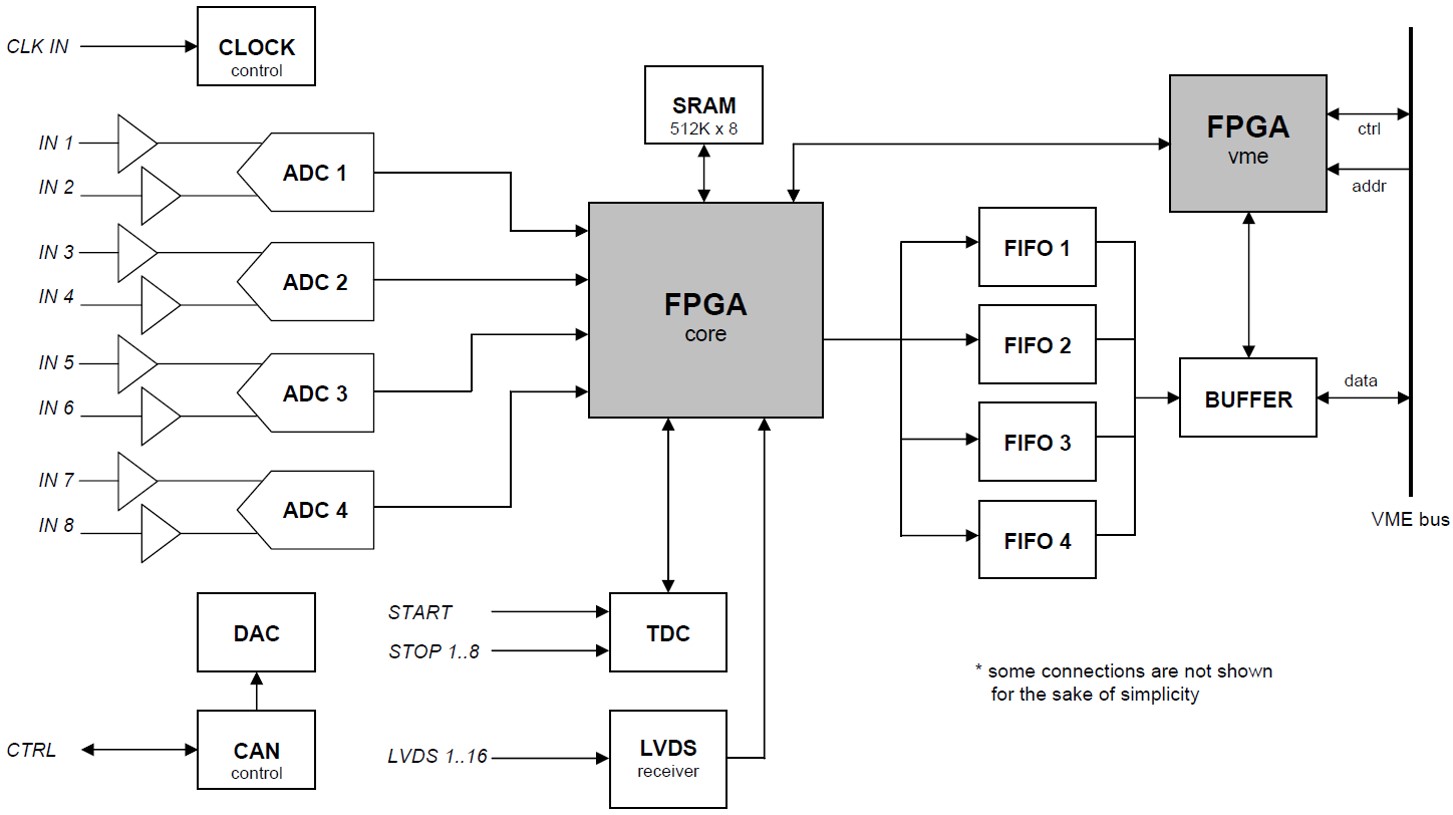}
\caption{Block diagram of the NDAQ data acquisition board.}
\label{fig:ndaq_diag}
\end{figure}

Each NDAQ module has eight Analog-to-Digital Converter (ADC) channels working at a 125~MHz sampling rate (10-bit resolution), as shown in Figure~\ref{fig:ndaq_diag}. The ADC output samples are sent to a Field Programmable Gate Array (FPGA), which controls the data flow to the two FIFO memories connected in series.

\subsection{Trigger}

When an event occurs, the output signals from the Front-End Electronics (FEE) are digitized by the NDAQ modules and stored in onboard FIFO (First-In, First-Out) memories pending a trigger decision. 
If the Trigger System selects the event~\cite{JMSSouza_M}, the data is then transferred to the DAQ server and subsequently recorded in the data storage unit.

The trigger board processes inputs from 40 discriminated signals, including 32 PMTs from the target, 4 PMTs from the lateral Veto, and 4 PMTs from the top Veto. 
An event is accepted if at least five PMT units from the Target are triggered within a specified time window, and it is blocked if simultaneous triggers are detected from two individual PMT units in either veto system.

\begin{figure}[ht!]
\centering
\includegraphics[scale=0.6]{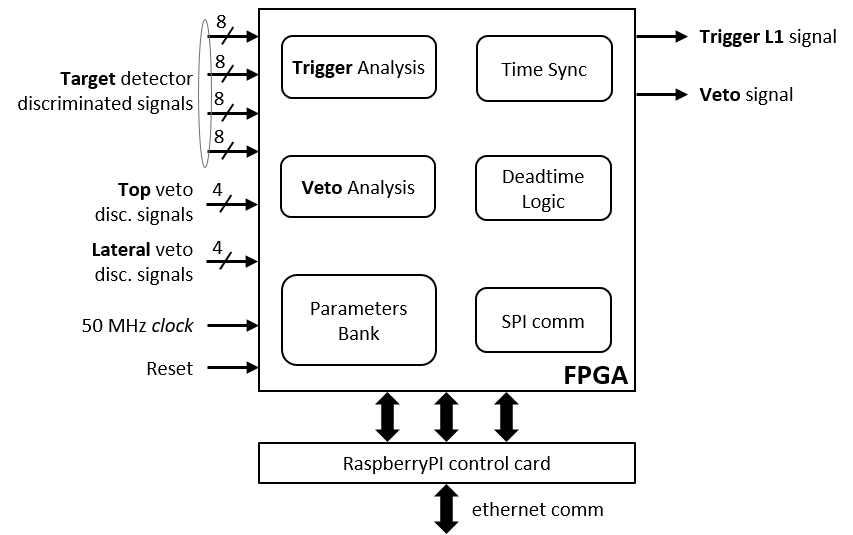}\\
\caption{Trigger electronics overview.}
\label{fig:trigger_diag}
\end{figure}

Different trigger configurations are available and can be configured by remote access to a Raspberry PI\textregistered control card. 

\subsection{Data flow control software}

To manage all data acquisition components, the detector utilizes a system orchestrator that runs a Python script known as Run-Control on the Angra DAQ Server. 
The primary role of Run-Control is to initiate data acquisition on the Readout Processor (ROP) and trigger system. 
Upon receiving data, a `Run' file is created and temporarily stored in RAM with a time stamp managed by the Redis network database. 
This `Run' record is then formatted into an Apache Parquet structured file. 
Subsequently, the raw file is transmitted to a Network Attached Storage (NAS) system over Gigabit Ethernet, which maintains a 4 hard drives, 4TB each, configured in RAID 5. 
Each `Run' captures 30 minutes of event data.

\begin{figure}[!ht] 
 \centering \includegraphics[width=0.7\columnwidth]{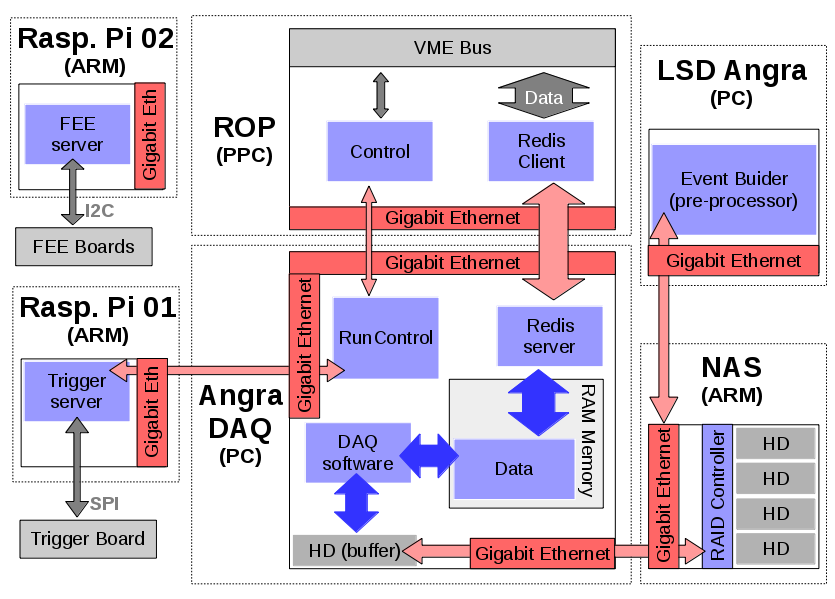}
 \caption{
   \label{fig:acquisitionData} 
   $\nu$-Angra Data Acquisition System diagram.
 }
\end{figure}

Pre-processing routines handle the data files to integrate the charge data for each PMT, enabling fast data analysis. 
These preprocessed files are also stored in the Apache Parquet format. 
Furthermore, a redundancy system mirrors these files on the CBPF and Unicamp clusters, making them accessible for data analysis to all $\nu$-Angra collaborators. 
A diagram of the data acquisition software setup is illustrated in Figure~\ref{fig:acquisitionData}.

\section{Analysis}

Our analysis pipeline comprises the following sequence of data handling: data quality assessment and preprocessing, energy calibration, event selection, and statistical analysis of reactor on/off cycles.
In the following subsections, we provide descriptions of each one of these steps.

\subsection{Data quality and pre-processing}

The data quality assessment and pre-processing phase has three main objectives:
\begin{itemize}
    \item To verify that the detector's physical operating conditions, including temperature and high voltage, remain within permissible limits and thus ensure the stability of the detector's performance;
    \item To quantify critical low-level physical parameters, such as PMT baselines, electronic noise, and the charge and timing of pulses.
    \item To calibrate the essential physical measurements, including the tuning of PMTs gain at the single photoelectron level. 
\end{itemize}

The fine adjustment of the PMT gains is achieved by aligning the peaks of single photoelectrons (SPEs). These SPEs originate from thermal noise and are identified via spectra in the very low dynamic range of our ADCs.  The calibration of the NUDAQ ADCs resulted in the following relationship between the number of photoelectrons and charge:

\begin{equation}
1 \text{ pe} = 78.0 \pm 0.6 \text{ DUQs}.
\label{eq:energiape2}
\end{equation}  

\begin{figure}[!h] 
 \centering \includegraphics[width=0.8\columnwidth]{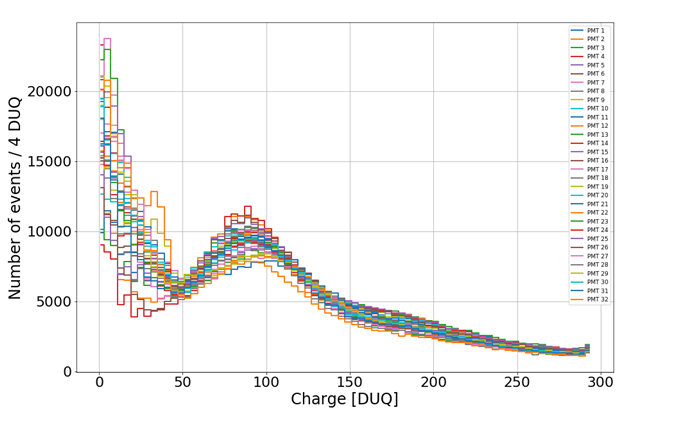}
 \caption{PMT gain equalization at the single photoelectron level. Each histogram shows the SPE spectrum of a single PMT.} \label{fig:photoelectron}
 \end{figure}

By individually adjusting the high voltage of each channel, it is possible to equalize the gains of the PMTs by alignment of all the SPE peaks around the same ADC value. The gain uniformity of the PMTs is shown in Figure \ref{fig:photoelectron}, which shows the charge distribution of PMT pulses at SPE levels, measured in Digital Units of Charge (DUQ).

\begin{figure}
 \centering \includegraphics[width=1.0\columnwidth]{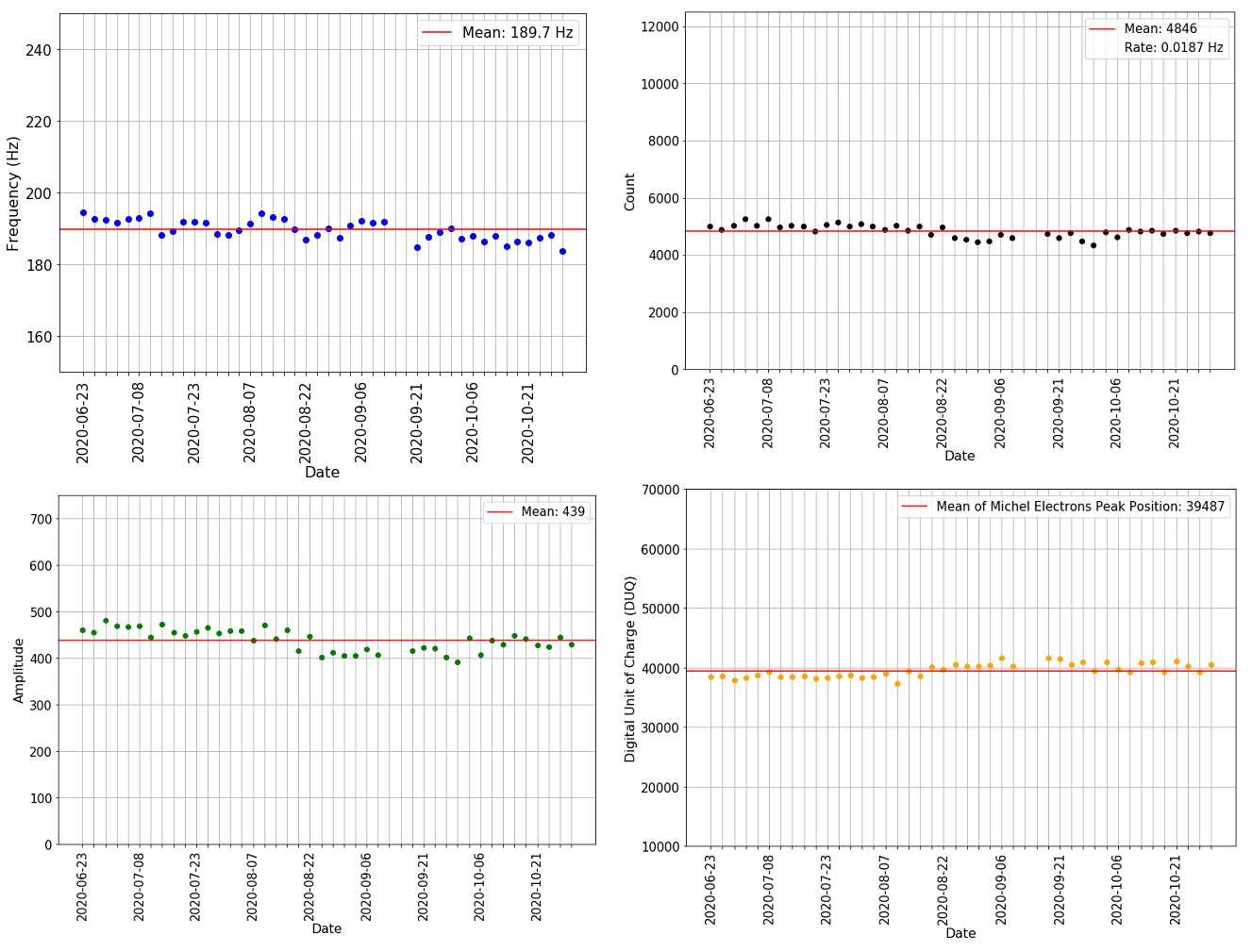}
 \caption{Upper left: Trigger rate of the $\nu$-Angra detector for the analyzed dataset; Upper right: Time evolution of Michel electrons (ME) counting; Lower left: Time evolution of ME spectrum amplitude; Lower right: Time evolution of ME spectrum peak position.}
\label{fig:triggee_em_rates} 
\end{figure}

The stability of the detector is crucial for the kind of analysis we are performing. Figure \ref{fig:triggee_em_rates} summarizes different indicators of detector stability for each three-day run during the second half of the 2020 data acquisition campaign, which corresponds to the data set examined in this study.

The upper left panel of Figure \ref{fig:triggee_em_rates} shows the trigger rate of the detector, as the first indicator of stability over the analyzed time frame. Another method of confirming the detector stability is through the analysis of Michel electrons (ME) from muon decays inside the target. Because this process is independent of the reactor operating state, the results of the ME analysis are a reliable and unbiased way to check the detector's sanity and stability. The ME selection criteria are described in detail elsewhere \cite{Gonzalez_D, Lima:2018spe}. To extract stability parameters from ME we use Gaussian fits on the ME spectra\footnote{A representative example of the shape of ME spectrum, as measured by the $\nu$-Angra detector, can be viewed in \cite{Lima:2018spe}.}. The upper right panel illustrates the consistent stability of the ME counting rate, while the lower left panel shows a similar trend for the ME peak amplitude, which is directly proportional to the ME rate. Moreover, the robustness of the energy measurements can be inferred from the plot in the lower right panel of Figure \ref{fig:triggee_em_rates}. This plot shows the position of the peak obtained from Gaussian fits over the ME spectra throughout the period of data analysis. The small fluctuations in the peak position imply stable energy measurements of the detector during this period. All of these results add confidence and reliability to our data analysis.

From the ME analysis, we also cross-checked our veto system efficiency. On average, the top veto tank has an efficiency of approximately 99.76\% \cite{Souza:2021kao}. The average ME rate obtained in our analysis is 0.02 Hz, representing 0.01\% of events in the total trigger rate, which falls within the 0.24\% of unvetoed events. This is a very consistent result, as not all unvetoed muons crossing the detector have energies in the range to be stopped inside the target.

\subsection{Calibration and event selection}

\subsubsection{Calibration}

\begin{figure}[!ht] 
 \centering \includegraphics[width=0.7\columnwidth]{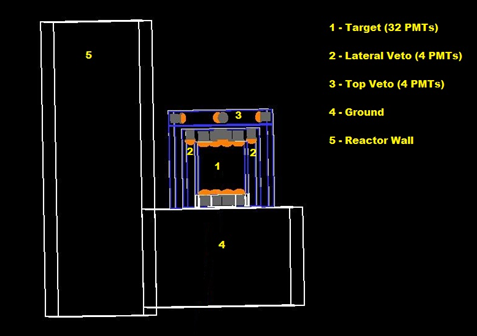}
 \caption{
   Cross-sectional view of the detector elements generated by our GEANT4 simulation code.
\label{fig:G4detView}
}
\end{figure}

To establish the relationship between collected charge and energy, our collaboration developed a comprehensive and detailed simulation of the detector using a GEANT4 code. Figure \ref{fig:G4detView} shows a cross-sectional view of the detector volumes (including the concrete wall of the reactor dome) of the simulation. The simulation also helps to determine the cut-off criteria for filtering background noise from the events recorded by the $\nu$-Angra detector.

\begin{figure}[!ht] 
 \centering \includegraphics[width=0.8\columnwidth]{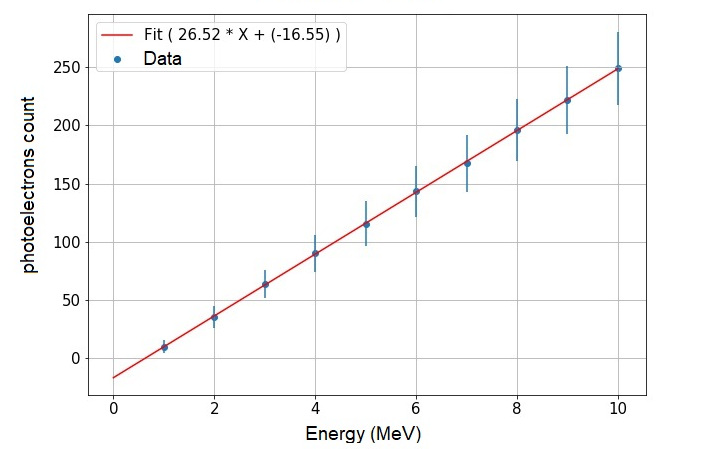}
 \caption{
   Calibration curve fit to obtain the photoelectron number $N_{pe}$ to energy conversion \cite{WVSantos_M}. Data are the Gaussian peaks from PE distributions obtained from simulations for each positron energy E. Error bars are the corresponding Gaussian widths.
\label{fig:calibrationFit}
}
\end{figure}

The energy calibration of our detector must be based on simulations due to stringent safety restrictions that prohibit the use and handling of radioactive sources within the nuclear power plant. To elucidate the correlation between energy and photoelectron (PE) yield, we conducted simulations of positrons at discrete energy values ranging from 1 MeV to 10 MeV, encompassing the energy range expected of positrons from IBD reactions induced by neutrinos from the reactor. For each specified energy value, 1200 positrons were fired with uniformly distributed random spatial positions and isotropic momentum vectors within the target volume of the detector. For each positron energy E, the PE distribution obtained from the simulation was fitted with a Gaussian, establishing the correlation between the positron energy E and the total number of PEs generated in the PMTs (the Gaussian peaks). Figure~\ref{fig:calibrationFit} illustrates the results clearly showing a linear relationship. From the linear fit, the conversion from the positron energy to the photoelectron count $N_{pe}$ is given by \cite{WVSantos_M}: 

\begin{equation}
N_{pe}(E) = 26.52 \cdot E - 16.55  
\label{eq:PE_x_E}
\end{equation}

Finally, one can convert the measured charge of the detector signals (in DUQ units) to energy (in MeV) by combining equations \ref{eq:energiape2} and \ref{eq:PE_x_E}.

\subsubsection{Event selection}\label{sec:criteria}

The IBD antineutrino interaction can be characterized by three properties related to the detector signals: prompt and delay energy (see section \ref{detector}), and the time interval between two signals.

To select events associated with IBD candidates, we rely on the energy calibration discussed above. The time interval is the only property not derived from the simulation; instead, it is determined directly from the $\nu$-Angra detector data, taking into account the Poissonian nature of the time between two events.

\paragraph{Event time interval:}

\begin{figure}[!ht] 
\begin{center}
\includegraphics[width=0.9\columnwidth]{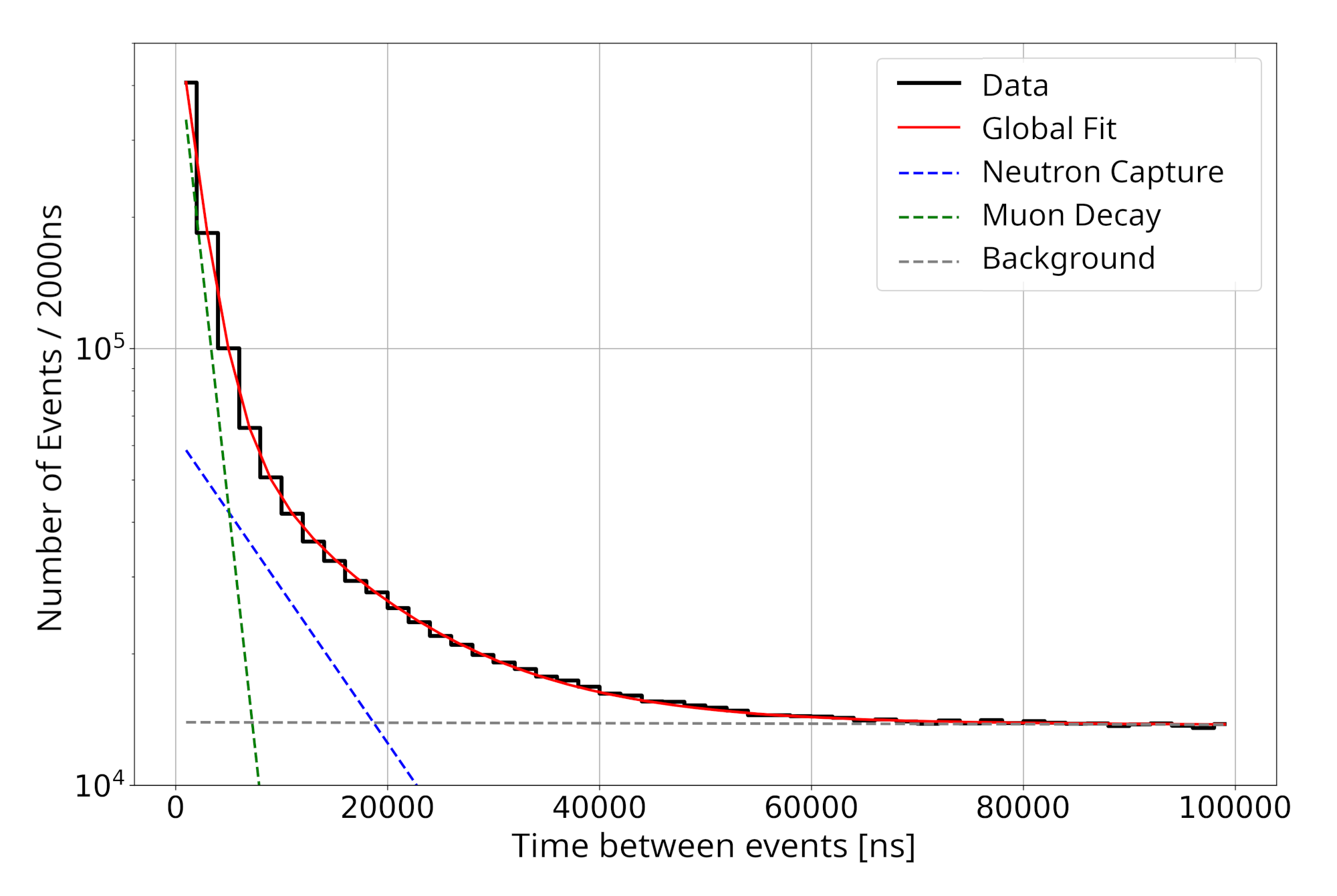}
 \caption{
   \label{fig:Poisson} 
   Time interval distribution obtained during the commissioning run of the $\nu$-Angra detector. The data (black histogram) was fitted with a sum of three components: muon decays, neutron captures and background. Red line is the overall fit, dotted lines are the fit results for each individual component. 
 }  
\end{center}
\end{figure}

Figure~\ref{fig:Poisson} displays the distribution of the time intervals between the events measured during the detector commissioning run \cite{Lima:2018spe}. From this distribution, namely a superposition of three Poissonian exponentials ($\propto exp(-t/\lambda)$), we have identified different physics processes: muon decay, neutron capture, and background noise.
Table ~\ref{tab:coefPoisson} lists the time constant $\lambda$ for each process, obtained from a fit of the data. 

\begin{table}[!htp]
\begin{tabular}{p{0.4\textwidth}|p{0.3\textwidth}}
\hline
\textbf{Component} & \textbf{$\lambda$}\\
\hline
Muon Decay & (1.9 $\pm$ 3) $\mu$s          \\
\hline
Neutron Capture & (12.32 $\pm$ 0.05) $\mu$s \\
\hline
Background Noise & (7.92 $\pm$ 0.01) ms \\ 
\hline
\end{tabular}
\caption{Time constants for each component of the time interval distribution. }
\label{tab:coefPoisson}
\end{table}

With muon decay occurring at $\sim$~2 $\mu$s and neutron capture at $\sim$~12 $\mu$s, we define a 4$\lambda$ range limit for these characteristic times. 
Consequently, we set the lower limit at 8 $\mu$s and the upper limit at 50 $\mu$s  in the time between two events as the initial criterion for event selection. Thus, the first and second events selected by this criterion define a pair (prompt,delay).

\paragraph{Prompt energy:}


%


The prompt energy should take into account the following expression $E_{e+} = E_{\Bar{\nu}} - \Delta$ MeV, where $E_{e+}$ is the positron energy, $E_{\Bar{\nu}}$ is the electron antineutrino energy from IBDs, and $\Delta = m_n - m_p = 1.293$ ~MeV, represents the mass difference between a neutron and a  proton.

\begin{figure}[!ht]
 \centering \includegraphics[width=.75\columnwidth]{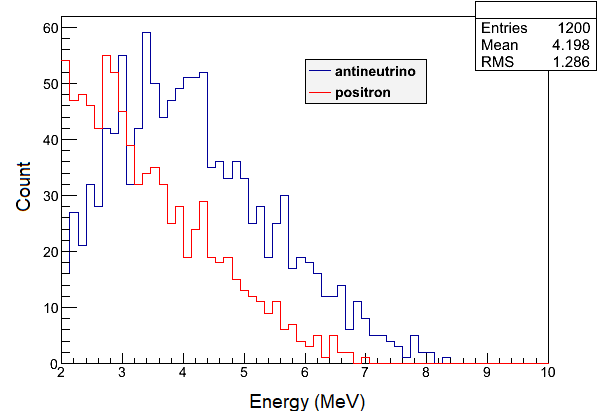}
 \caption{
   \label{fig:Sorteio} 
   Positron drawn energy from reactor antineutrino spectrum.
}
\end{figure}

The energy cut-off points for prompt events was determined from simulations of the detector response for IBD positrons. Antineutrino energies are drawn from a PDF given by the normalized convolution of the spectrum of $\beta$-decays in the reactor's fuel the ($\Bar{
\nu_e}$,p) cross-section, then converted to positron energy using Equation \ref{eq:epositron}. These energies are then assigned to positrons with uniformly distributed positions and isotropic momenta within the target volume. The resulting antineutrino and positron spectra used to simulate IBD prompt events in the detector are shown in Figure \ref{fig:Sorteio}.

\begin{figure}
\centering \includegraphics[width=0.7\columnwidth]{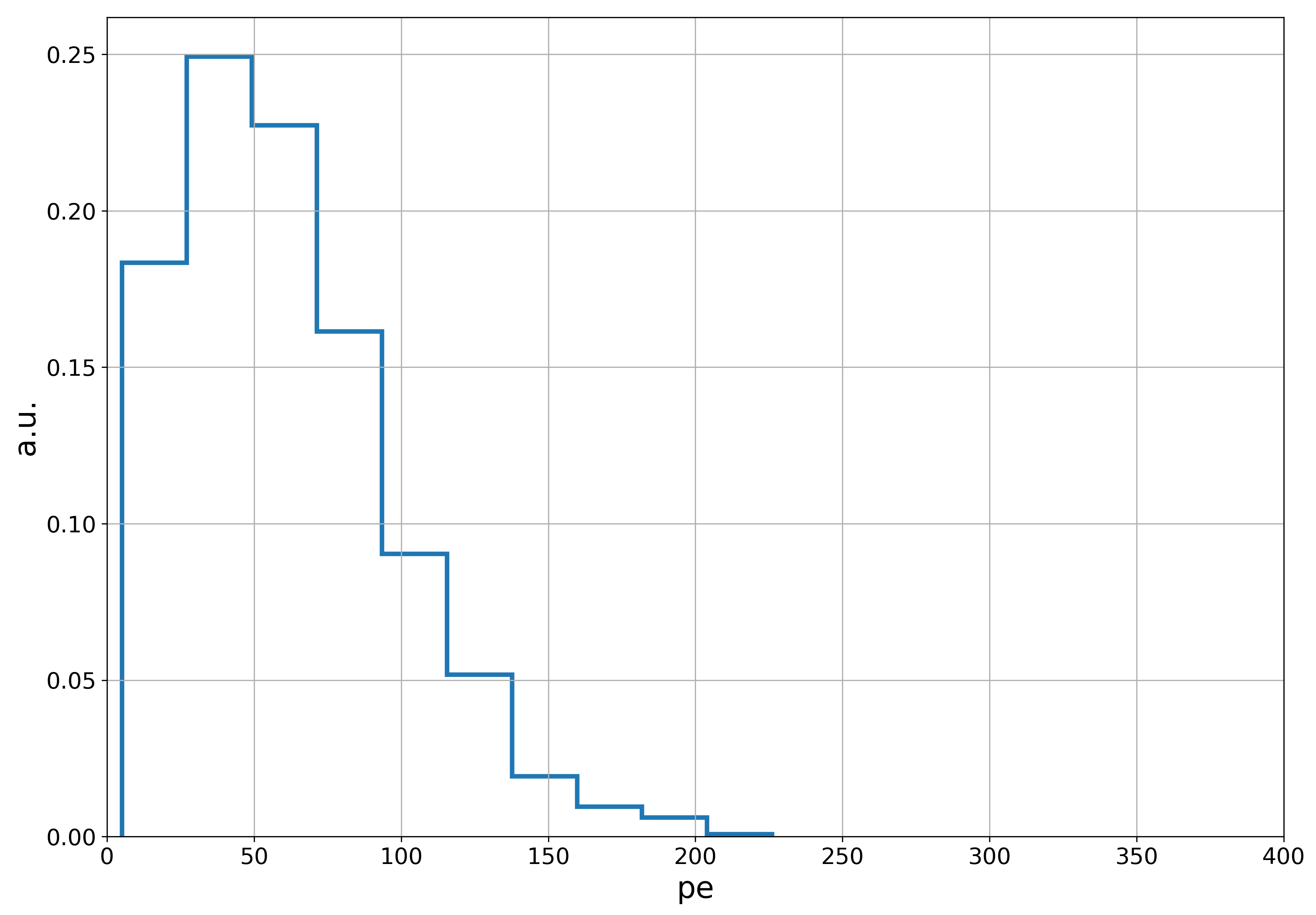}
\caption{Detector response for IBD positrons generated from reactor antineutrino spectrum (prompt signal PDF).}
\label{fig:distPrompt} 
\end{figure}

The IBD positrons generate Cherenkov photons that are collected by the detector's PMTs, resulting in the normalized distribution shown in Figure~\ref{fig:distPrompt}.  This simulation followed the minimum trigger requirement, which is the activation of a minimum of five PMTs.
However, the lower-energy region is heavily influenced by background noise from sources such as gammas from cosmic rays and external neutrons. 
To filter out these background events and tag the primary event of the pair, we only accept events with prompt energies $E_p$ in the range of 3 MeV $\le E_p \le$ 10 MeV.

\paragraph{Delay energy: }

\begin{figure}[!ht] 
\centering \includegraphics[width=0.8\columnwidth]{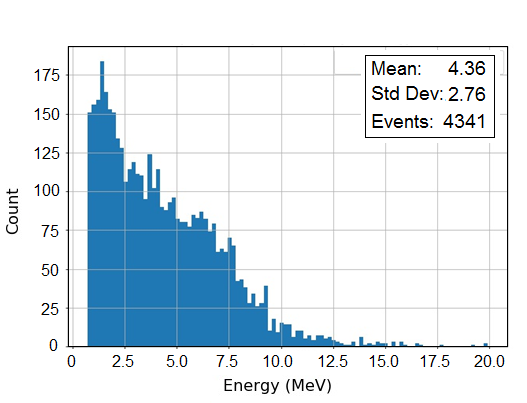}
 \caption{Expected spectrum of the delay signals, produced by gammas from the de-excitation of Gd after neutron capture \cite{WVSantos_M}.}
   \label{fig:Neutrons} 
\end{figure}

\begin{figure}
\centering \includegraphics[width=.7\columnwidth]
 {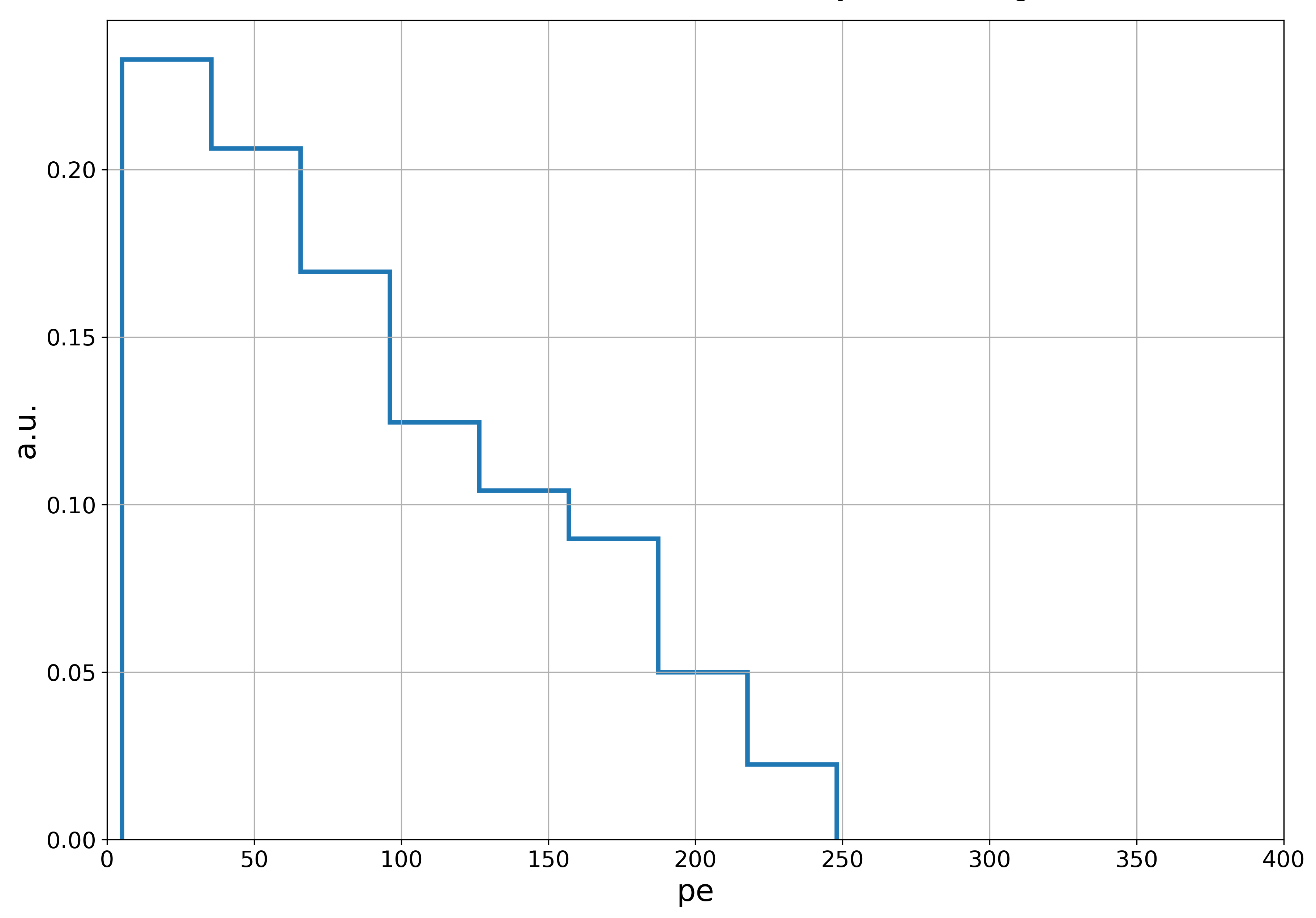}
 \caption{Detector response for the gammas from de-excitation of Gd atoms in the target after neutron capture (delay signal PDF).}
\label{fig:distDelay}  
\end{figure}

In addition to the filtering based on time interval and prompt energy, the selection criterion for the event pair includes the delayed signal. This third cut-off criterion was established by simulating neutrons with an energy of 10 keV, with their momentum and position uniformly randomized within the target. The selection of low neutron energy is a natural outcome of IBD kinematics and is also essential to achieve rapid thermalization in the simulation, ensuring neutron capture by gadolinium. Neutron capture is followed by the emission of deexcitation gammas $\gamma_{Gd}$ that generate Cherenkov radiation through Compton scattering with electrons in the medium. The spectrum of gammas of the reaction n(Gd, Gd$^*$)$\gamma$ is shown in the histogram of Figure~\ref{fig:Neutrons}, and after normalization we have the PDF of Figure~\ref{fig:distDelay}, with probabilities in terms of the number of PE generated in the detector PMTs.

Using the mean energy and RMS of the $\gamma_{Gd}$ expected spectrum (delay signals) we can set the lower and upper energy limits for the delay energy E$_d$ cutoff as 1.60 MeV $\le E_d \le$ 7.12 MeV.

\begin{figure}[!ht] 
 \centering \includegraphics[width=0.7\columnwidth]{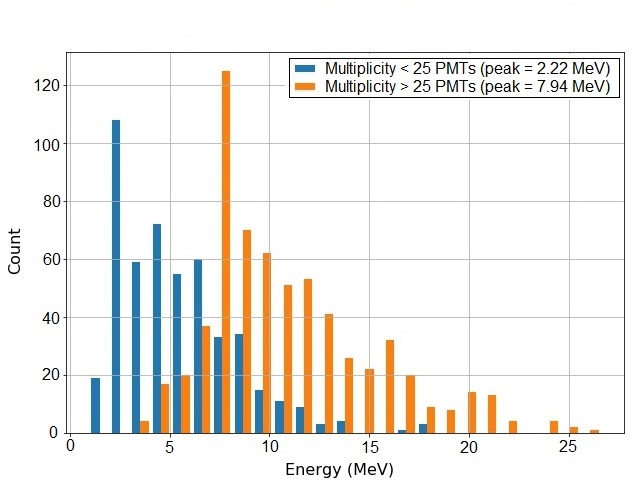}
 \caption{Simulated spectra of gammas from neutron capture filtered by the multiplicity M of PMTs activated in the events. M$\ge$25 selects efficiently gammas Gd de-excitation from those of deuteron formation.}
\label{fig:Regimes} 
\end{figure}

Another additional criterion derived from the simulations is the PMT multiplicity in each event. The simulation results analysis revealed that the $\gamma_{Gd}$ of approximately 8 MeV activate more than 25 PMTs. Signals with fewer than 25 PMT were generally due to hydrogen capture $n+p\rightarrow D+\gamma$, which releases typically a gamma with energy of 2.2 MeV after deuteron deexcitation, as illustrated in Figure~\ref{fig:Regimes}.

\subsection{ON-OFF analysis}\label{sec:result}

The most straightforward way to check the detector ability to monitor reactor activity is the ON-OFF analysis. In this analysis, reactor neutrino IBD events are identified by comparing datasets collected when the reactor is operational (ON) and during reactor shutdowns for maintenance and refueling (OFF). During the ON periods, the reactor produces a flux of antineutrinos, absent during the OFF periods, which serves as a background reference. By comparing the event rates and characteristics between the ON and OFF datasets, we can isolate and confirm the presence of reactor antineutrino IBD events, effectively distinguishing them from background noise and other sources of interference.

The current analysis was performed over 3 data sets around the reactor shutdown in 2020. The reactor operation periods are from 08/20/2020 to 09/13/2020, and from 09/24/2020 to 09/28/2020, totalizing a first block of 30 ON days (ON1), and a second one from 10/01/2020 to 10/30/2020 also with 30 ON-days (ON2). These two blocks of ON data were compared to a non-operational period from 06/24/2020 to 07/23/2020 (30 OFF days). We applied the event selection criteria outlined in Section~\ref{sec:criteria} on these data sets to select pairs of delayed events corresponding to candidates for IBD. The prompt spectra of both the ON and OFF periods were obtained and then subtracted from each other. 

\begin{figure}[!ht] 
 \centering \includegraphics[width=0.8\columnwidth]{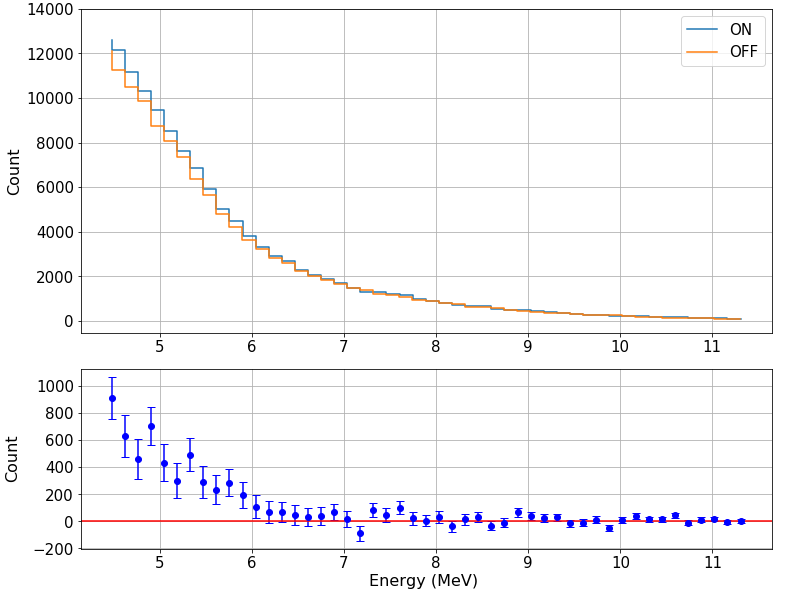}
 \caption{
   \label{fig:ONOFF} 
   ON1 and OFF prompt spectra (upper panel). The ON1-OFF subtraction, shows a clear excess of the reactor-ON period  (lower panel).  
 }
\end{figure}

The results of the ON1 data set against the OFF period are shown in Figure~\ref{fig:ONOFF}.

An observable excess in the 4.5 - 6.5 MeV range is completely compatible with the expected energy spectrum of antineutrino events from IBD. A $\chi^2$ hypothesis test was conducted assuming as null hypothesis $H_0$ the OFF spectrum, implying that any excess observed could be due to statistical background fluctuations. 

\begin{figure}[!ht] 
 \centering \includegraphics[width=0.8\columnwidth]{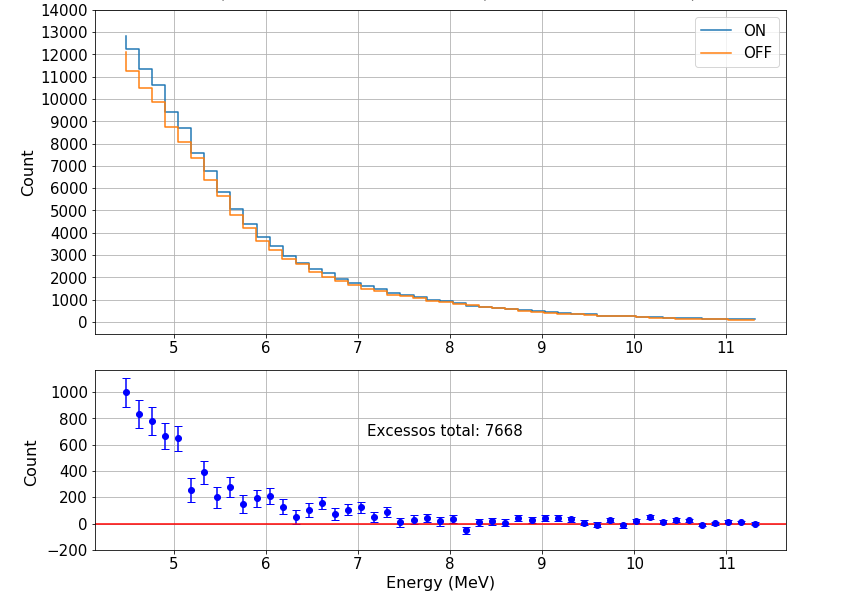}
 \caption{
   \label{fig:ONOFF2} 
   ON2 and OFF prompt spectra (upper panel). The ON2-OFF subtraction, shows a clear excess of the reactor-ON period  (lower panel)
 }
\end{figure}

In our analysis, the null hypothesis $H_0$ was decisively rejected by $\chi^2/dof \sim 13.4$, which implies an almost zero p-value. The same test was applied to the subsequent 30-day ON2 period, with results shown in Figure \ref{fig:ONOFF2}. In this case, $\chi^2/dof \sim 17.3$ further reinforced the rejection of $H_0$, thus confirming the previous ON1-OFF results as a genuine excess.

\section{Conclusions}

The $\nu$-Angra experiment aims to demonstrate significant advancements in the field of neutrino detection technology through the deployment of a water Cherenkov detector at the Angra dos Reis nuclear power plant. 
This detector may be capable of monitoring antineutrino emissions from nuclear reactors, thus enhancing nuclear safety measures.

We successfully implemented a sophisticated data acquisition and analysis system for the $\nu$-Angra experiment. This system played a crucial role in guaranteeing the quality of the data taken from 2018 to 2022.

Our calibration processes, using experimental data and simulations, were crucial in refining the detector's sensitivity to prompt and delayed neutrino signals.

The application of GEANT4 simulation to define the selection criteria for the IBD candidates has enabled us to accurately differentiate between neutrino interactions and background noise. 

The ON/OFF analysis conducted during different operational phases of the reactor provided hints of the detector's capability to distinguish between reactor operational states, which is pivotal for reactor monitoring and safeguards. We clearly observed the excess of IBD candidates from the reactor-ON periods of the 2020 detector runs. The statistical significance of the excess is remarkable. However, recognition as genuine IBD events depends on a deeper understanding of the background. Future efforts will aim to enhance our event selection criteria by thoroughly examining the background noise, which is essential for accurate identification of IBD events.
Our results demonstrate the potential of the $\nu$-Angra detector technology for non-proliferation purposes. We should highlight that we encountered the difficult task of operating a water Cherenkov neutrino detector at surface level. Nevertheless we achieved significant background suppression by integrating the detector's hardware capabilities with suitable data analysis techniques, aligning with the International Atomic Energy Agency's strategies to enhance nuclear safeguards tools. The ability to monitor reactors in near-field scenarios with different technologies enhances transparency and compliance with international nuclear nonproliferation treaties.\\
\\
\\

\paragraph{\large Acknowledgements\\    }

We express our heartfelt gratitude to the late Dr. Ademarlaudo F. Barbosa, without whom this experiment would never have been possible. We
hope we have made you proud.\\
\\
\\
\\


\bibliographystyle{unsrt}
\bibliography{angra-onoff-arxiv}

\end{document}